\documentclass[aps,prc,twocolumn,tightenlines,showpacs,
superscriptaddress,nofootinbib,14pt]{revtex4}
\usepackage[figuresright]{rotating}

\begin{document}

%\preprint{}

\title{Existence of long-lived isotopes of a superheavy element
\\
 in natural  Au }


\author{A. Marinov}
\email{marinov@vms.huji.ac.il} \altaffiliation{Fax:
+972-2-6586347.}
 \affiliation{Racah Institute of Physics, The Hebrew
University of Jerusalem, Jerusalem 91904, Israel}
\author{I. Rodushkin}
%\thanks{}
\affiliation{Analytica AB, Aurorum 10, S-977 75 Lule\aa, Sweden}
\author{A. Pape}
\affiliation{IPHC-UMR7178, IN2P3-CNRS/ULP, BP 28, F-67037
Strasbourg cedex 2, France}
\author{Y. Kashiv}
\affiliation{Racah Institute of Physics, The Hebrew University of
Jerusalem, Jerusalem 91904, Israel}
\author{D. Kolb}
\affiliation{Department of Physics, University GH Kassel, 34109
Kassel, Germany}
\author{R. Brandt}
\affiliation{Kernchemie, Philipps University, 35041 Marburg,
Germany}
\author{R. V. Gentry}
\affiliation{Earth Science Associates, P.O. Box 12067, Knoxville,
TN 37912-0067, USA}
\author{H. W. Miller}
\affiliation{P.O.Box 1092, Boulder, CO 80306-1092, USA}
\author{L. Halicz}
\affiliation{Geological Survey of Israel, 30 Malkhei Israel St.,
Jerusalem 95501, Israel}
\author{I. Segal}
\affiliation{Geological Survey of Israel, 30 Malkhei Israel St.,
Jerusalem 95501, Israel}


\date{February 20, 2007}
%\date\today

%\keywords{Superheavy elements; Alpha decay; Proton decay; Isomeric
%states; Superdeformation; Hyperdeformation}
%\keywords{}
\begin{abstract}
Evidence for the existence of long-lived isotopes with atomic mass
numbers 261 and 265 and abundance of (1-10)x10$^{-10}$ relative to
Au has been found in a study of natural Au using an inductively
coupled plasma - sector field mass spectrometer. The measured
masses  fit  the predictions made for the masses of  $^{261}$Rg
and $^{265}$Rg (Z=111) and for some isotopes of nearby elements.
 The possibility that these isotopes
belong to the recently discovered class of long-lived high spin
super- and hyperdeformed isomeric states is discussed.
\end{abstract}

\maketitle

%\section{Introduction}
Long-lived isomeric states of $^{211}$Th, $^{213}$Th, $^{217}$Th
and $^{218}$Th have been reported recently  in natural Th,  with
abundances of (1-10)x10$^{-11}$ relative to $^{232}$Th
\cite{mar06}. The evidence for the existence of the isomeric
states in the Th nuclei was obtained in accurate mass measurements
using an inductively coupled plasma - sector field mass
spectrometer (ICP-SFMS), operated at medium mass resolution. These
results motivated us to search for a similar phenomenon in the
region of superheavy elements. If, for instance, Rg (Z=111) exists
in nature, it might be found with Au, its nearest chemical
homologue. Unlike Th, where the g.s. masses of the
neutron-deficient isotopes are known \cite{aud03}, in the case of
the relevant superheavy elements, one has to rely on mass
predictions. One finds, however, that various predictions
\cite{mol95,kou03,lir01} for the atomic masses of the Rg isotopes
with mass numbers between 259 and 269 (mass region studied here)
differ from one another by at the most 0.003 u. This value is
small compared to the experimental uncertainty in our ICP-SFMS
measurements. Also, the average prediction
\cite{mol95,kou03,lir01} for the mass of the g.s. of for instance
$^{261}$Rg is 261.154 u and for the g.s. mass of $^{261}$Au
\cite{mol95,kou03} is 261.331 u, values well separated by  the ICP
- SFMS.

In principle, native Au would be the best material to analyze.
However, it was found \cite{mar06} that background was the main
obstacle  when looking for isotopes with relative abundances  of
(1-10)x10$^{-11}$. Therefore, pure natural Au solutions were used
in our measurements. It is reasonable to assume that Rg (if it
exists), or a significant part of it, will follow  Au in the
chemical purification process. In the present work, we have
performed accurate mass measurements for masses 254 (for checking
and correcting the calibration using the $^{238}$U$^{16}$O peak)
and 259 to 269. Evidence for the existence of isotopes with masses
that fit the predictions for the masses of $^{261}$Rg and
$^{265}$Rg was obtained.

The experimental procedure was similar to that described earlier
\cite{mar06}. The instrument was an  Element2 (Finnigan,
Thermo-Electron, Bremen, Germany). The predefined medium
resolution mode m/$\Delta$m = 4000 (10\% valley definition) was
used throughout the experiments to separate atomic ions from
molecular interferences with the same mass number. The
sensitivity-enhanced setup of the instrument was similar to that
described in Ref. \cite{rod04}. This setup provided
 sensitivity for $^{197}$Au in this resolution mode of up to 7x10$^{7}$ counts
  s$^{-1}$mg$^{-1}$l$^{-1}$. The sample uptake rate was approximately 60-80 $\mu$l
  min$^{-1}$.
  Methane gas was added to the plasma to decrease the formation of
  molecular ions \cite{rod05}. Oxide and hydride formation (monitored
  as UO$^{+}$/U$^{+}$
  and UH$^{+}$/U$^{+}$ intensity ratios) were approximately 0.04 and 1x10$^{-5}$,
   respectively.
   Mass calibration was performed  using the $^{115}$In$^{+}$,
   $^{232}$Th$^{+}$, $^{235}$U$^{+}$,
   $^{238}$U$^{+}$ and $^{238}$U$^{16}$O$^{+}$ peaks. The $^{115}$In peak was used
   for on-line mass drift correction. Two stock solutions,
   ``A" and ``B", of 1000 mg Au in 1 l of 10\% HCl,
    were obtained from
   two manufacturers (Customer Grade and Spex,
    LGC Promochem AB, Bor\aa s, Sweden). Complete elemental screening was
    performed on both solutions to assess the impurity concentration
     levels. The following concentrations
     (expressed as ppm
      of the Au concentration) of certain trace elements that can potentially give rise
      to spectrally interfering molecular species are given below:

A: U 0.01, Th 0.3, Bi 67, Pb 11, Tl 0.1, Hg 0.3, Pt 104, Os 16, Gd
0.005, Ba 0.9, Cs 0.05, I 10, Te 0.07, Ag 0.4, Zn 3, S 69, Si 51,
Al 19, Na 33.

B: U 0.01, Th 0.3, Bi 104, Pb 18, Tl 0.3, Hg 1, Pt 124, Os 25, Gd
0.005, Ba 0.9, Cs 0.06, I 5, Te 0.1, Ag 0.6, Zn 6, S 182, Si 237,
Al 36, Na 149.

             \begin{figure}[h]
\vspace*{-0.5cm} \hspace*{-0.2cm}
\includegraphics[width=0.5\textwidth]{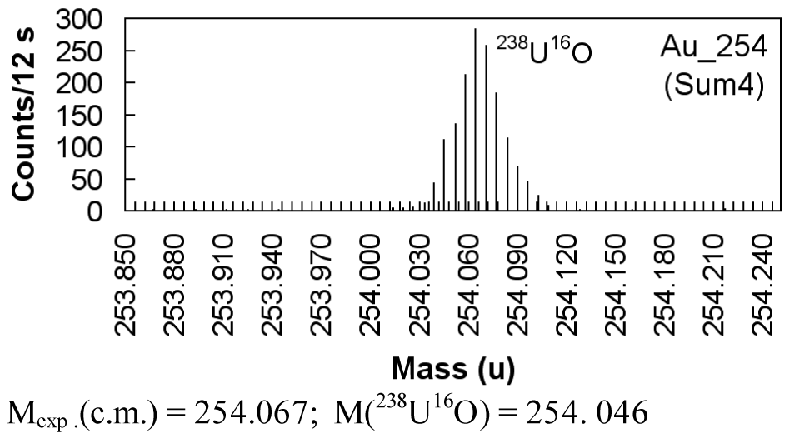}
%\vspace*{-0.2cm}
 \caption{Measurements for mass region 254
 obtained with the Au solutions.
 The sum of four measurements, two with
 solution A
 and two with solution B, is displayed. M$_{exp.}$(c.m.) - observed mass
 position;  M($^{238}$U$^{16}$O) -
 known mass of $^{238}$U$^{16}$O
  taken from Ref. \cite{aud03}.}
\end{figure}
%\vspace*{0.5cm}

 The solutions were analyzed  on September 13, 2006.
  A range of about 0.4 u, divided into 65 channels, was scanned in each
 measured spectrum.
 The  masses 254 and 259 to 269 were analyzed
 with an integration time per channel of 3 s.
   Both solution A and solution B were diluted to 20 mg Au l$^{-1}$ of
    0.7 M HNO$_{3}$, and each was
      measured twice. Altogether, four spectra were
    taken with the Au solutions for each mass number studied.  Replicate
    analyses of blank solution (0.7 M HNO$_{3}$) were performed. Instrumental sensitivity
  varied  among runs as a result of matrix effects caused by
  the introduction of highly concentrated solutions into the ICP source.

Figure 1 shows the sum spectrum of the four measurements of mass
 254, where the peak of  $^{238}$U$^{16}$O$^{+}$ is
seen. The FWHM of the peak is about 0.030 u. The spectrum
   shows
a shift of 0.021 u in the peak position. A similar shift, of 0.015
u was observed at mass 265 for the
$^{232}$Th$^{16}$O$_{2}$$^{1}$H$^{+}$ peak (Fig. 3(b)).  Based on
these shifts, a correction of -0.018 u was applied throughout the
measurements. However, shifts of up to about 0.025 u
     were seen in some of the measurements.

                          \begin{figure}[h]
%\vspace*{-0.5cm}
\hspace*{-0.2cm}
\includegraphics[width=0.5\textwidth]{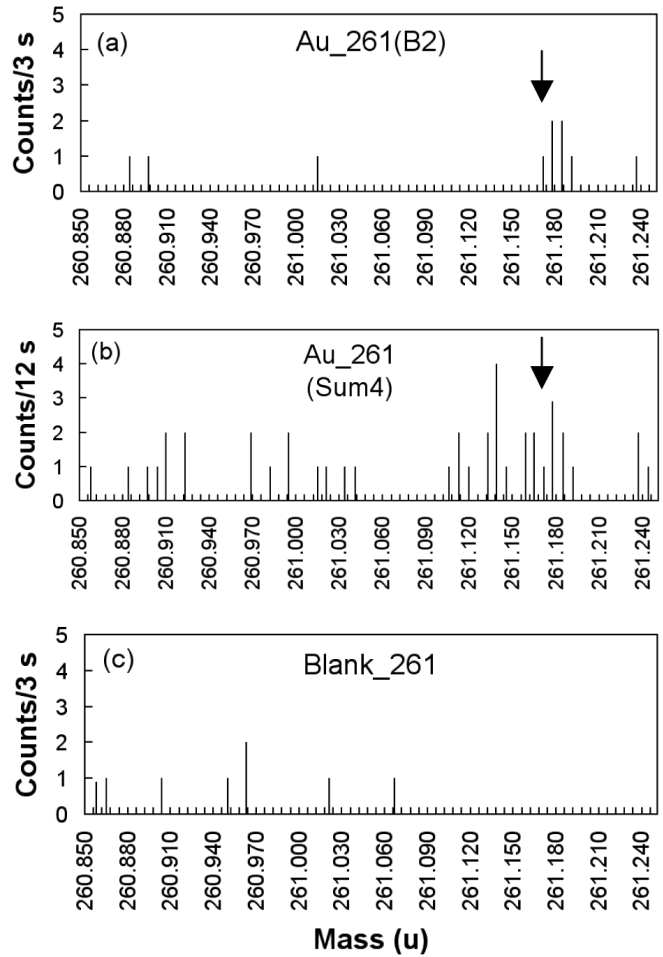}
%\vspace*{-0.2cm}
 \caption{Results of measurements for mass region 261. Figure 2(a) shows
 the results obtained in the second measurement with solution B.
  The sum of four
 spectra, two with
 solution A
 and two with solution B, is displayed in Figure 2(b).
 A spectrum of a blank solution in given in Figure 2(c).
 The arrows indicate
 the position of the predicted $^{261}$Rg mass (see text).}
\end{figure}
%\vspace*{0.5cm}

    Below are the results for masses 261 and 265.
     For each mass number, three spectra are presented. The first is
     the ``best" of  the four measured spectra. The second is
     the sum of the four individual spectra, and the third shows
     a spectrum obtained with the blank.

                   \begin{figure}[h]
%\vspace*{-0.5cm}
\hspace*{-0.2cm}
\includegraphics[width=0.5\textwidth]{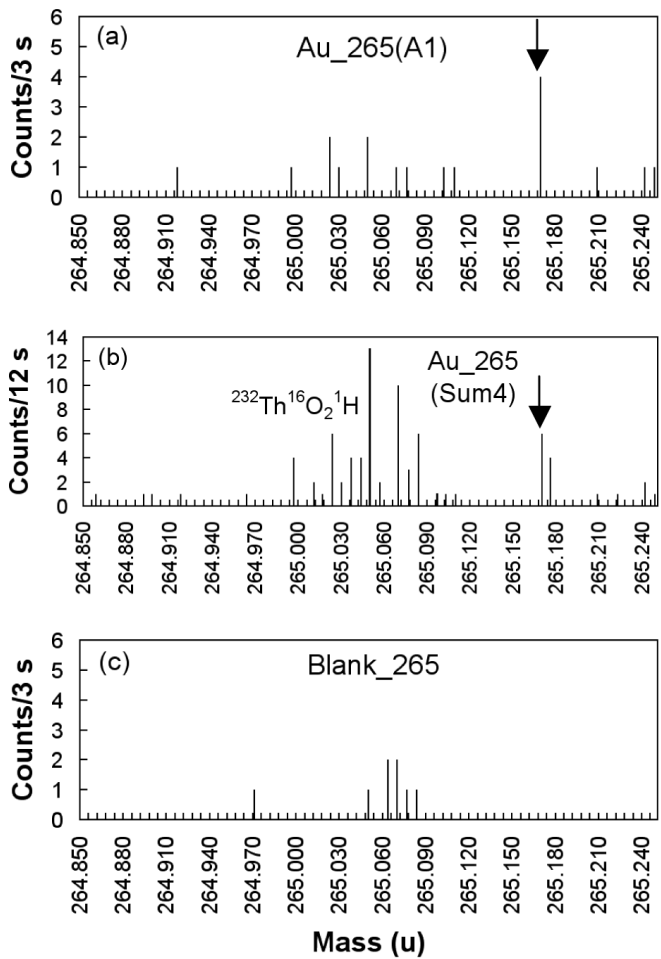}
\vspace*{-0.6cm}
 \caption{Results of measurements for mass region 265. Figure 3(a) shows
 the results obtained in the first measurement with solution A.
  The sum of four
 spectra two with
 solution A
 and two with solution B, is displayed in Figure 3(b).
 A blank solution spectrum in given in Figure 2(c).  The arrows indicate
 the position of the predicted $^{265}$Rg mass (see text).}
\end{figure}
%\end{figure}
%\vspace*{0.5cm}

 Figure 2 displays the results for mass 261. The arrows in Fig. 2(a) and 2(b)
 indicate the position of the  predicted  $^{261}$Rg mass, corrected for the
 shift discussed above. This mass was
 taken as the average of
 the predictions given in
 Refs. \cite{mol95,kou03,lir01}.
  A pronounced group of
 six events is seen In Fig. 2(a), at a mass that fits the $^{261}$Rg mass prediction.
 In Fig. 2(b) a group of 22 events is seen at a mass that fits
  the predicted mass
 of $^{261}$Rg. Although this group is somewhat broad, we assume
 that it is a single group since it is not broader than the
 $^{232}$Th$^{16}$O$_{2}$$^{1}$H group seen in Fig. 3(b).
 (The width of the latter is typical for a single group, since, as can be
 seen in Table II,
 the masses of the Th and U molecules are very close to one
 another, and all the other possible molecules have lower masses.)
 However, if one assumes that the four events at the low mass end of this peak
 are due to background, then 18 events are seen in this group.

  The results  obtained for mass 265 are given in Fig. 3.
    A group of
    four events is seen in Fig. 3(a)  at a mass that fits the predicted mass of $^{265}$Rg.
    The sum spectrum  of the four individual measurements is presented in Fig. 3(b).
        In addition to a
    peak due to  $^{232}$Th$^{16}$O$_{2}$$^{1}$H, a group of ten events
     is seen at a mass
     which fits the prediction for the mass of  $^{265}$Rg.

     The statistical significance of the newly observed peaks, that
correspond to the predicted masses of the corresponding Rg
isotopes, was calculated as described in Ref. \cite{mar06}: The
probability P'$_{acc.}$ that n out of a total number of N
           randomly distributed counts will occur accidentally in a
           small mass region r out of a total
 mass region R
           is given by

\vspace*{0.1cm}
 P'$_{acc.}$ = $\mbox{$(^{N}_{n})$
          (r/R)$^{n}$(1-(r/R))$^{(N-n)}$}$.\hspace{2.0cm}(1)

 \vspace*{0.1cm}
 The total probability that such a group will occur
           accidentally in a predefined region r out of a total
           region R is given by

\vspace*{0.1cm}
 P$_{acc.}$ = P'$_{acc.}$(r/R). \hspace{4.4 cm}(2)

\vspace*{0.1cm}
 In calculating the
values of P$_{acc.}$ for the data given in Figs. 2(a) and 2(b),
the value of R was chosen as the whole measured region. For the
results displayed in Figs. 3(a) and 3(b), the region of the
molecular ion $^{232}$Th$^{16}$O$_{2}$$^{1}$H was removed from the
total region.

                %\begin{widetext}

%\end{widetext}
      %\begin{widetext}

        \begin{table}[h]
\centering
\caption{Summary of results of mass measurements and
comparison with the predicted masses of $^{261}$Rg and
$^{265}$Rg.}
%\vspace*{-0.3cm}
%\begin{minipage}{\textwidth}
\begin{minipage}{0.5\textwidth} % Embed the whole table in a 'minipage' for sake of
                                % displaying table footenotes (if any)
\renewcommand{\footnoterule}{\kern -3pt} % Needed for omitting the separation rule
%\begin{center}
%\begin{center}
\begin{tabular}{llclcc}
\hline Mass & Fig. &  No. of  & \hspace{0.2cm}P$_{acc.}$
&\hspace{0.2cm}
 M$_{c.m.}^{exp.}$\footnote{The uncertainty in mass is estimated to be $\pm$0.025 u.}
  & Mass of\\
 no. & no. &  events &   &
 &\hspace{0.2cm} Rg isotope\footnote{Average of predicted
 values, Refs. \cite{mol95,kou03,lir01}.}\\ \hline
\\[-10pt]
261 & 2(a) & 6 & 5x10$^{-7}$&  &  \\
261 & 2(b) & 22(18) & 3x10$^{-6}$\footnote{Because of the
different widths of the lines, the same value is obtained for 22
and 18 events lines.}&\hspace{0.2cm} 261.134\footnote{For 18
counts
M$_{c.m.}^{exp.}$=261.142} & 261.154 \\
265 &3(a) & 4 & 3x10$^{-7}$&   &  \\
265 & 3(b) & 10 & 1x10$^{-9}$&\hspace{0.2cm} 265.154 & 265.151 \\
\hline
\end{tabular}
\end{minipage}
\renewcommand{\footnoterule}{\kern-3pt \hrule width .4\columnwidth
\kern 2.6pt}            % Recreation of the footnote separation rule (if any)
%\end{center}
\vspace{-0.3cm}
\end{table}

%\end{widetext}
%\vspace{-0.6cm}

      The results are summarized in Table I.
Column 3 gives  the  number of
        events, with measured masses  in accord with
         the predicted
        masses of the corresponding Rg isotopes.
         P$_{acc.}$ values are given  in column 4.
           Column 5 gives
        the measured
        masses of the  events in Figs. 2(b) and 3(b), corrected for the 0.018 u shift
        in mass calibration.
        Column 6 gives
        the averages of the predictions in Refs. \cite{mol95,kou03,lir01}
        for the g.s. masses of the Rg isotopes with the same
        mass number.
        It should be mentioned that the different predictions differ
         by no more than 0.002 u and 0.003 u  from one another for
        $^{261}$Rg and $^{265}$Rg, respectively.
\vspace{-0.6cm}
        \begin{table}[h]
\centering \caption{Known masses \cite{aud03} of several ions with
molecular masses of 261 and 265, and the predicted masses of
$^{261}$Au \cite{mol95,kou03} and $^{265}$Au \cite{kou03}.}
%\vspace*{-0.3cm}
%\begin{minipage}{\textwidth}
\begin{minipage}{0.5\textwidth} % Embed the whole table in a 'minipage' for sake of
                                % displaying table footenotes (if any)
\renewcommand{\footnoterule}{\kern -3pt} % Needed for omitting the separation rule
%\begin{center}
%\begin{center}
\begin{tabular}{cccc}
\hline
 &\hspace{-1.5cm} 261 &   & \hspace{-1.8cm}265 \\
Molecular & Mass & Molecular &Mass   \\
%\vspace*{-0.2cm}
ion&              & ion&             \\
 \hline
\\[-10pt]
$^{238}$U$^{23}$Na &\hspace{0.2cm}261.041
&\hspace{0.2cm}$^{238}$U$^{27}$Al &
\hspace{0.2cm}265.032  \\
$^{232}$Th$^{29}$Si &\hspace{0.2cm}261.015
&\hspace{0.2cm}$^{232}$Th$^{16}$O$_{2}$$^{1}$H &
\hspace{0.2cm}265.036  \\
$^{197}$Au$^{64}$Zn &\hspace{0.2cm}260.896
&\hspace{0.2cm}$^{197}$Au$^{68}$Zn &
\hspace{0.2cm}264.891  \\
$^{133}$Cs$^{128}$Te &\hspace{0.2cm}260.810
&\hspace{0.2cm}$^{138}$Ba$^{127}$I &
\hspace{0.2cm}264.810  \\
$^{261}$Au &\hspace{0.2cm}261.331 &\hspace{0.2cm}$^{265}$Au &
\hspace{0.2cm}265.360  \\
\hline
\end{tabular}
\end{minipage}
\renewcommand{\footnoterule}{\kern-3pt \hrule width .4\columnwidth
\kern 2.6pt}            % Recreation of the footnote separation rule (if any)
%\end{center}
%\vspace{-0.3cm}
\end{table}

%\end{widetext}

\vspace*{-0.4cm}

           We have been unable to
              match the  signals of the suspected Rg isotopes
              with any molecular ions.
               Usually, as
              seen in Table II and in Figs. 1 and 3(b),
              because of the binding energies,
               the masses of the molecules are
              lower than the masses of the new peaks reported
              here.  On the other hand, the predicted
              masses  of the extremely neutron-rich nuclei
              $^{261}$Au and $^{265}$Au are higher than the
              observed peaks.%\vspace*{-0.5cm}
              \footnote{In Fig. 2b,
              some events are seen in
              the mass regions of $^{232}$Th$^{29}$Si
              and $^{197}$Au$^{64}$Zn. However, since we did not
              see the $^{232}$Th$^{28}$Si peak at
              mass 260 ($^{28}$Si/$^{29}$Si $\simeq$ 20),
              the events around 261.015 u are
              not due to $^{232}$Th$^{29}$Si. As mentioned,
                 P$_{acc.}$  for the peak at 261.134 u was estimated
                  assuming that all the events outside
               this peak are due to background.}

                             Another
                possibility that should be considered is the potential presence of
                hydrocarbon-based molecular ions from pump oil. However, there are no
                hydrocarbon candidates for masses 261 and 265, and typical
                hydrocarbon masses lie well above  the predicted
                masses
                of the Rg isotopes with the same mass number. For instance, the masses
                of CH$_{3}$(CH$_{2}$)$_{16}$CH$_{3}$
                and CH$_{3}$(CH$_{2}$)$_{17}$CH$_{3}$ are
                 254.297 u and 268.313 u \cite{aud03}, respectively,
                 whereas the predicted masses of the respective Rg isotopes
                  are 254.164 u \cite{kou03} and 268.152 u \cite{mol95,kou03,lir01}.
On the other hand, it is evident from the data  in Table I that
the
        masses of the  measured  new peaks fit predictions for the g.s.
        masses \cite{mol95,kou03,lir01}
        of the corresponding Rg isotopes within 0.020 u.

        Based on the
         mass measurements alone, one cannot conclude whether
         the newly observed peaks are due to Rg isotopes or
        to isotopes of the adjacent
        superheavy elements Ds and element 112. Since the
        peaks appeared in practically  pure Au solutions, where the concentration of
        Pt and Hg, the nearest chemical homologues of Ds and element
        112, are about 1x10$^{-4}$ and 1x10$^{-6}$ of
        Au, respectively (see above), it is reasonable to assume that they could be due to
        Rg.

          It is estimated that the concentration of these
                  Rg isotopes amounts to (1-10)x10$^{-10}$ of
                  $^{197}$Au
                  (about (2-20)x10$^{-15}$
of the solutions).

  The predicted g.s. half-lives of these Rg isotopes  are
      very short, of the order of  1 $\mu$s \cite{mol97}.
      This suggests  that the observed events are due to
          long-lived Rg isomers. (The  accuracy of the
      present experiment is
      not sufficient  to determine the excitation energies of the isomeric
states relative  to the predicted normal g.s. masses.) If
       their initial terrestrial concentration was similar to that of $^{197}$Au,
       then the lower limit on their half-lives would be about 1x10$^{8}$ y.

      The character of the observed isomeric states has not
      been measured directly. As discussed in the case of the Th isomers
      (Ref. \cite{mar06} and references therein), it seems that these isomeric states
       cannot
       represent the high spin isomers that occur near doubly closed shells,
    since they are far from the nearest predicted \cite{nil69,mos69} spherical doubly
         closed shells at Z = 114 and N = 184. Nor can they be related to
       the fission isomers found in actinides, since their lifetimes are
       in the region of ns to ms. As mentioned in Ref. \cite{mar06}, it is also
       not reasonable to assume that they are due to  high spin K-type
        isomers, since the lifetimes of all the known K-isomeric
           states in neutron-deficient nuclei with Z$\geq$84,
          are not longer than several minutes. It might be reasonable to assume
         that as in the case of the Th isomers, the isomeric
          states seen in the present experiment are due to the
          recently discovered  aligned high spin
          superdeformed (SD) and/or hyperdeformed (HD)
           states  \cite{mar96a,mar96b,mar01a,mar01b},
          where the high spin, the barriers between the various minima of the
          potential energy surfaces and the unusual radioactive decay properties
contribute to their long lifetimes.

       High spin states in general and such states
      in  SD and HD minima in particular are preferentially
      produced by heavy ion reactions \cite{mar01b,mar03a}. The
      observation of these isomeric
      states  suggests
      that heavy ion reactions could be involved in their nucleosynthesis
     production process(es).

       In summary, evidence for the existence of long-lived superheavy isotopes
               with t$_{1/2}$$\geq$1x10$^{8}$ y  and atomic numbers of 261 and 265
                has been found in natural
        Au. It is argued that they are isomeric states, most probably in
        $^{261}$Rg and $^{265}$Rg.
         The possibility that they are high spin SD and HD  isomeric states
         is discussed.

       We appreciate valuable discussions and the help of N. Zeldes,
       E. Grushka,
        and  O. Marinov.

\end{document}